\documentclass[a4paper,fleqn,usenatbib,useAMS]{mnras}

\usepackage{mathptmx}

\usepackage[T1]{fontenc}
\usepackage{ae,aecompl}

\usepackage{graphicx}	
\usepackage{amsmath}	
\usepackage{amssymb}	

\newcommand{\bz}{\langle B_z \rangle}

\newcommand{\kms}{km\,s$^{-1}$}

\title[The two magnetic components of HD\,104237]{
The two magnetic components in the Herbig~Ae SB2 system HD\,104237}

\author[S. P. J\"arvinen et al.]{
S.~P.~J\"arvinen$^{1}$\thanks{E-mail: sjarvinen@aip.de},
T.~A.~Carroll$^{1}$,
S.~Hubrig$^{1}$,
I.~Ilyin$^{1}$,
M.~Sch\"oller$^{2}$,
\and
N.~A.~Drake$^{3,4}$,
M.~A.~Pogodin$^{5}$
\\
$^{1}$Leibniz-Institut f\"ur Astrophysik Potsdam (AIP), An der Sternwarte~16, 14482~Potsdam, Germany\\
$^{2}$European Southern Observatory, Karl-Schwarzschild-Str.~2, 85748 Garching, Germany\\
$^{3}$Laboratory of Observational Astrophysics, Saint Petersburg State University, Universitetsky~pr.~28, 198504~Saint~Petersburg, Russia\\
$^{4}$Laborat\'orio Nacional de Astrof\'{\i}sica /MCTIC, Rua Estados Unidos, 154, CEP 37504-364, Itajub\'a, MG, Brazil\\
$^{5}$Central Astronomical Observatory at Pulkovo, Pulkovskoye chaussee~65, 196140~Saint~Petersburg, Russia
}

\date{Accepted 2019 XXX. Received 2019 YYY; in original form 2019 ZZZ}

\pubyear{2019}

\begin{document}
\label{firstpage}
\pagerange{\pageref{firstpage}--\pageref{lastpage}}
\maketitle

\begin{abstract}
  We present longitudinal magnetic field measurements $\bz$ for the Herbig\,Ae
  primary and the T\,Tauri secondary in the SB2 system HD\,104237. These
  measurements were carried out using high spectral resolution observations
  obtained with the High Accuracy Radial velocity Planet Searcher in
  polarimetric mode, installed at the ESO La Silla 3.6\,m telescope. In
  agreement with previous studies of Herbig\,Ae stars, the longitudinal
  magnetic field in the primary is rather weak, ranging from 47\,G to 72\,G.
  The secondary component possesses a variable, much stronger magnetic field,
  up to 600\,G, as expected for a magnetically active T\,Tauri star. We
  estimated the rotation period of the primary,
  $P_{\mathrm{rot}}=4.33717\pm0.00316$\,d ($=104\pm0.08$\,h), from metal line
  equivalent width variations.
  \end{abstract}




\section{Introduction}
\label{sec:intro}

It is generally accepted that magnetic fields are important ingredients in the
star formation process
\citep[e.g.][]{McKee}
and are already present in stars during the pre-main sequence (PMS) phase.
However, it is still not clear whether these fields persist until the main
sequence stage in the intermediate mass Herbig~Ae stars with radiative
envelopes. Studies of magnetic fields in such stars are of special interest to
get an insight into the origin of strong kiloGauss-order magnetic fields
observed in Ap and Bp stars on the main sequence.

While T\,Tauri stars with strong magnetic fields stand out by their strong
emission in chromospheric and transition region lines, the presence of weak
magnetic fields in the higher mass Herbig~Ae/Be stars has long been suspected.
The model of magnetically driven accretion and outflows successfully
reproduce many observational properties of the classical T\,Tauri stars, but
this picture is completely unclear for the Herbig~Ae/Be stars due to the poor
knowledge of their magnetic field topology. So far, only about 20 Herbig
stars have been reported to host magnetic fields
\citep[][and references therein]{Hubrig2015},
and the magnetic field geometry has been constrained only for two Herbig~Ae/Be
stars, V380\,Ori
\citep{alecian2009}
and HD\,101412
\citep{Hubrig2011}.
Magnetic fields in Herbig~Ae stars are generally very weak: only a few stars
have magnetic fields stronger than 200\,G and half of the known cases
possesses magnetic fields of about 100\,G or less
\citep{Hubrig2015}.

HD\,104237 (DX\,Cha) is a spectroscopic and visual binary system with a
companion at a distance of $2.2\pm0.7$\,mas
\citep{garcia2013}.
The primary component is known to show $\delta$\,Scuti-like pulsations
\citep{boehm2004}.
\citet{garcia2013}
give mass estimates for both companions, $M_1=2.2\pm0.2\,M_\odot$ and
$M_2=1.4\pm0.3\,M_\odot$, and the system inclination angle
$i=17\degr^{+12}_{-9}$.
\citet{boehm2004}
estimated the orbital elements of the system $P_{\rm orb}=19.859$\,d and
$e=0.665$. The rotation period $P_{\mathrm{rot}}=100\pm5$\,h of the primary has
been reported by
\citet{boehm2006}.

The basic stellar parameters of HD\,104237 were determined in a number of
studies
\citep{Grady2004,boehm2004,Fumel2012,Cowley2013}.
The most recent analysis by
\citet{Cowley2013}
yielded $T_{\rm eff}=8250$\,K, $\log g=4.2$, and $v\,\sin\,i= 8$\,km\,s$^{-1}$
for the primary and $T_{\rm eff} = 4800$\,K, $\log g=3.7$, and
$v\,\sin\,i= 12$\,km\,s$^{-1}$ for the secondary. The spectrointerferometric
study by
\citet{garcia2013}
with AMBER on the Very Large Telescope Interferometer in the K-band continuum
and the Br$\gamma$ line suggested the presence of a circumbinary disc with a
radius of about 0.5\,AU. However, about 50~per cent of the flux remained
unresolved and not fully accounted for by the stellar photospheres. The
authors suggested that this unresolved flux likely arises in compact
structures inside the tidally disrupted circumbinary disk.

The possible presence of a magnetic field in HD\,104237 of the order of 50\,G
was announced over 20 years ago by
\citet{Donati1997}.
However, the first low-resolution polarimetric spectra
with FORS\,1 on the Very Large Telescope yielded a non-detection
\citep{Wade2007}.
The observations obtained with the University College London {\'E}chelle
Spectrograph (UCLES)/Semelpol at the Anglo-Australian Telescope (AAT) showed
a field of negative polarity, but no field strength was reported
\citep{Wade2011}.
\citet{Hubrig2013}
estimated $\bz =63\pm15$\,G for the primary from a HARPSpol spectrum obtained
on May 3, 2010. More recently,
\citet{silva2015}
reanalyzed the 2010 data using a different technique and reported a definite
detection for the T\,Tauri secondary ($\bz =129\pm12$\,G) and a marginal
detection of 13\,G for the primary. 

In the following, we report on our most recent longitudinal magnetic field
measurements in both components of this SB2 system using ESO archival
observations obtained with the High Accuracy Radial velocity Planet Searcher
polarimeter
\citep[HARPS\-pol;][]{snik2008}
in March 2015.


\section{Observations and magnetic field measurements}
\label{sec:obs}

\begin{figure}
 \centering 
        \includegraphics[width=\columnwidth]{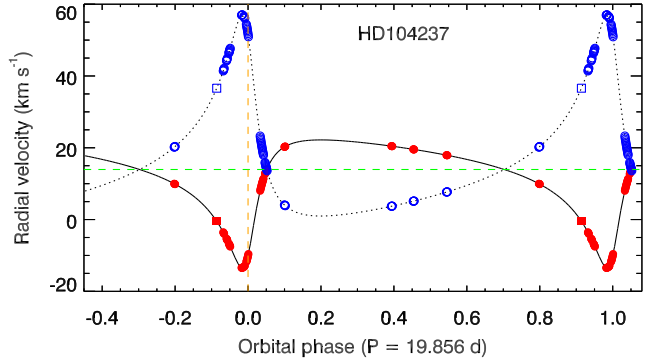}
        \caption{
          The orbital phases of the observations, following the orbital
          parameters of
          \citet{boehm2004}.
          The filled symbols correspond to the primary component and the
          open symbols to the secondary. The square symbol at phase 0.914
          shows the observation published by
          \citet{silva2015}.
          The horizontal dashed line indicates the $\gamma$ velocity and
          the vertical dashed line indicates $\varphi=0$ (periastron).
        }
   \label{fig:PhaseBoem}
\end{figure}

All HARPS spectropolarimetric observations used in our study have a spectral
resolution of about 115\,000 and cover the spectral range 3780--6910\,\AA{},
with a small gap between 5259\,\AA{} and 5337\,\AA{}. Each observation
obtained in 2010 consisted of eight subexposures with exposure times of about
two minutes whereas in 2015 each observation consisted of four subexposures with
exposure times of about four minutes. The quarter-wave retarder plate was
rotated by $90\degr$ after each subexposure. The final polarimetric
spectrum is the combination of the subexposures recorded at four different
positions of the quarter-wave retarder plate. The reduction and calibration of
these spectra was performed using the HARPS data reduction software available
at the ESO headquarter in Germany. The continuum normalization of the spectra 
is described in detail by 
\citet{Hubrig2013}.
The distribution of the observations over the orbital phase is illustrated in
Fig.~\ref{fig:PhaseBoem}. The orbital solution of this system was presented
by
\citet{boehm2004}:
${\rm HJD}=2451647.505$, $P_{\rm orb}=19.859$\,d, $e=0.665$,
$\gamma=13.943$\,\kms, $K_{1}=17.8$\,\kms.

\begin{figure*}
 \centering 
 \includegraphics[width=\textwidth]{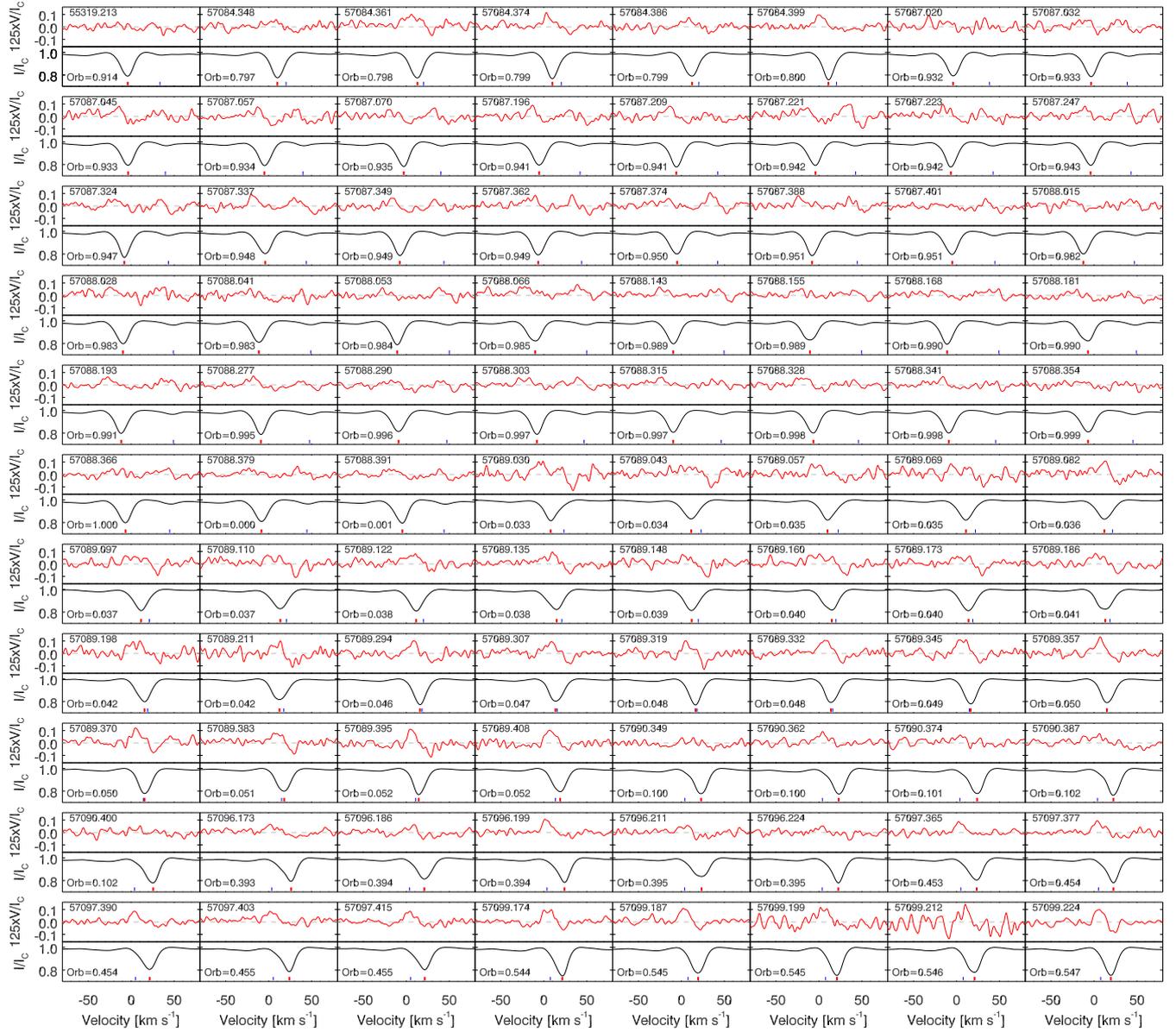}
        \caption{
          Stokes~$I$ (bottom panel) and $V$ (top panel)
          Least-Square-Deconvolution profiles for all 88 observation. The
          Stokes~$V$ profiles have been amplified by a factor of 125. The
          primary is marked with thick red and the secondary with thin blue
          ticks below the Stokes~$I$ profiles.
        }
   \label{fig:LSDIV}
\end{figure*}

As all HARPS spectra have a rather low signal-to-noise ratio (S/N), in the
range of 60--100, to increase the S/N, we applied the Least Squares
Deconvolution
\citep[LSD;][]{Donati1997}
technique. LSD is a cross-correlation technique for computing average Stokes
profiles from tens or hundreds of spectral lines simultaneously. It is based
on the assumption that all spectral lines have the same profile and that they
can be added linearly. Our line mask containing 715 spectral lines made use of
the Vienna Atomic Line Database
\citep[VALD; e.g.][]{Kupka2011,VALD3}
and was based on the stellar parameters of the primary component
of HD\,104237
\citep[$T_{\rm eff}=8250$\,K, $\log g=4.2$;][]{Cowley2013}.
The results of the application of the LSD technique to all available 88
spectra are presented in Fig.~\ref{fig:LSDIV}.

As illustrated in this figure, the LSD spectra still appear very noisy.
Therefore, to more accurately characterize the magnetic field variability in
both components, we applied the dedicated Singular Value Decomposition
technique
\citep[SVD;][]{Carroll2012}
to a number of spectra recorded around three orbital phases, which represent
the distribution of the observations over the orbital cycle. Around phase 0.94
we see the largest separation of the components and around orbital phases
0.45 and 0.54 the components come closer and appear blended. The SVD approach
is very similar to that of the Principle Component Analysis (PCA). In this
technique, the similarity of the individual Stokes~$V$ profiles allows one to
describe the most coherent and systematic features present in all spectral
line profiles as a projection onto a small number of eigenprofiles. The
excellent potential of the SVD method, especially in the analysis of weak
magnetic fields in Herbig~Ae stars, was already presented by
\citet{Hubrig2015}
and
\citet{silva2015,aksco}.

\begin{figure}
 \centering 
        \includegraphics[width=\columnwidth]{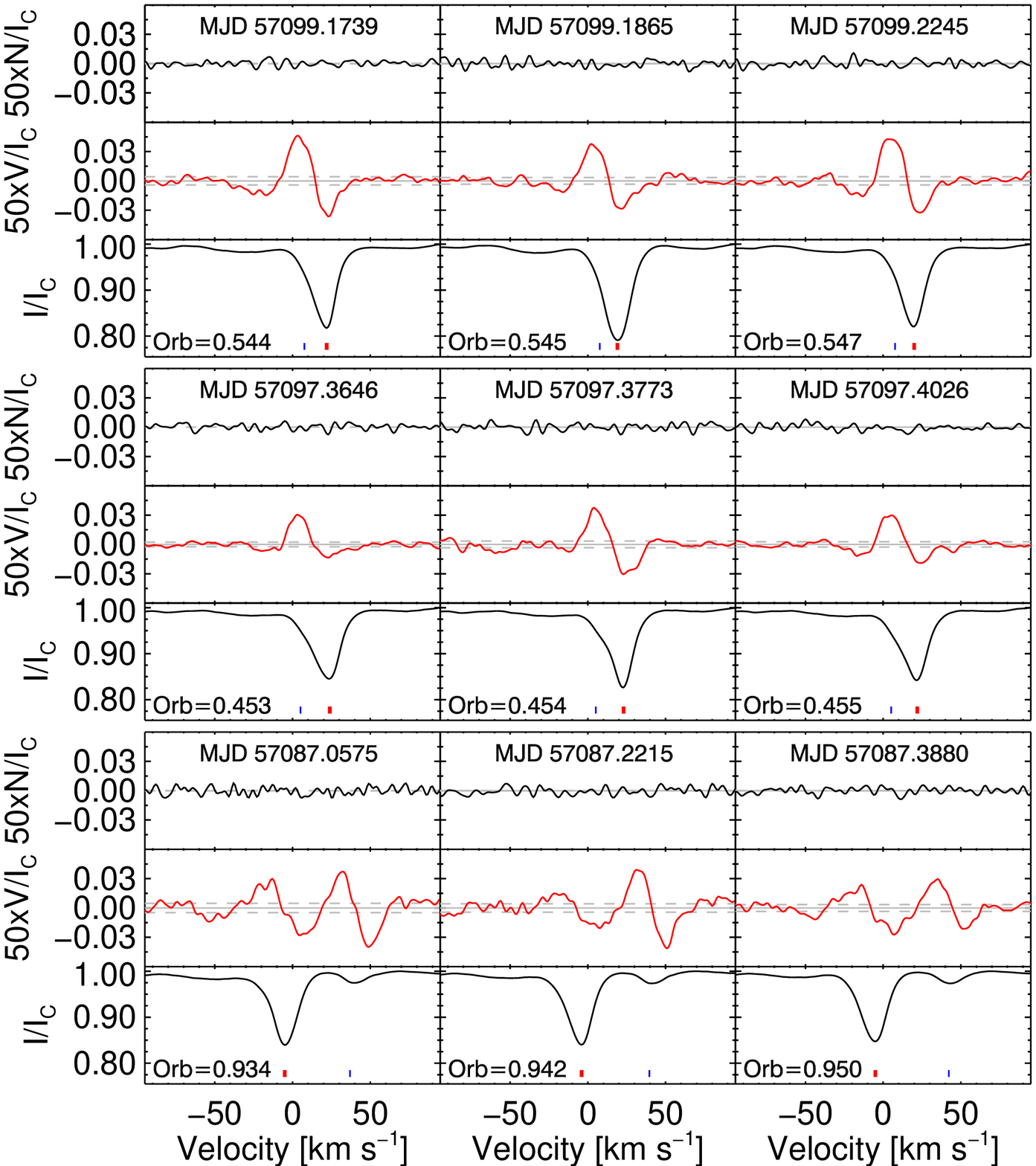}
        \caption{
          Single-Value-Decomposition Stokes-$I$ (bottom), $V$ (middle), and
          diagnostic null (N) profiles (top) obtained for HD\,104237 at
          different orbital phases. The Stokes~$V$ and $N$ profiles have been
          amplified by a factor of 50. The profiles are sorted according to
          the observing date. The components in the system are marked with
          ticks below the Stokes~$I$ profiles. The horizontal dashed lines
          indicate the $\pm1\sigma$-ranges.
        }
   \label{fig:IVN_HD104237}
\end{figure}

\begin{table}
\centering
\caption{
  Mean longitudinal magnetic field strengths of HD\,104237 on different
  orbital phases obtained using the Single-Value-Decomposition method. The
  columns present the heliocentric Julian date of mid-exposure followed by the
  orbital phase, the magnetic field strength for the Herbig\,Ae component
  (prim) and the corresponding rotation phase (see Sect.~\ref{sec:Prot}), and
  the magnetic field strength for the T\,Tauri secondary (sec).}
\label{tab:mfield}
\begin{tabular}{cc r@{$\pm$}l c r@{$\pm$}l }
\hline
HJD  & $\varphi_{\textrm{orb}}$ &
\multicolumn{2}{c}{$\left<B_{\rm z}\right>_{\rm prim}$} &
$\phi_{\textrm{rot}}$ & 
\multicolumn{2}{c}{$\left<B_{\rm z}\right>_{\rm sec}$} \\
2\,400\,000+ & &
\multicolumn{2}{c}{(G)} &
& 
\multicolumn{2}{c}{(G)} \\
\hline
57097.8646 & 0.453 & \multicolumn{2}{c}{---} & 0.666 & 144 & 15 \\
57097.8773 & 0.454 & \multicolumn{2}{c}{---} & 0.669 & 410 & 22 \\
57097.9026 & 0.455 & \multicolumn{2}{c}{---} & 0.675 & 189 & 19 \\
57099.6739 & 0.544 & \multicolumn{2}{c}{---} & 0.083 & 366 & 21 \\
57099.6865 & 0.545 & \multicolumn{2}{c}{---} & 0.086 & 349 & 22 \\
57099.7245 & 0.547 & \multicolumn{2}{c}{---} & 0.095 & 338 & 20 \\
55319.2133$^{1}$ & 0.914 & 13 & 8 & 0.687 & 129 & 12 \\
57087.5575 & 0.934 & 72  & 6  & 0.290 & 609 & 27 \\
57087.7215 & 0.943 & 47  & 6  & 0.328 & 440 & 23 \\
57087.8880 & 0.950 & 63  & 6  & 0.366 & 124 & 13 \\
\hline
\end{tabular}
\begin{minipage}{\textwidth}
{\bf Note:} $^1$ From \citet{silva2015}.
\end{minipage}
\end{table}

The mean longitudinal magnetic field is deduced by computing the first-order
moment of the Stokes~$V$ profile according to
\citet[][]{Mathys1989}:

\begin{equation}
\left<B_{\mathrm z}\right> = -2.14 \times 10^{11}\frac{\int \upsilon V
  (\upsilon){\mathrm d}\upsilon }{\lambda_{0}g_{0}c\int
  [I_{c}-I(\upsilon )]{\mathrm d}\upsilon},
\end{equation}

\noindent
where $\upsilon$ is the Doppler velocity in \kms, and $\lambda_{0}$ and
$g_{0}$ are the mean values for the wavelength (in nm) and the Land\'e factor
obtained from all lines used to compute the SVD profile, respectively. The
results of the magnetic field measurements are presented in
Table~\ref{tab:mfield} and the corresponding SVD profiles are shown in
Fig.~\ref{fig:IVN_HD104237}.

The three SVD profiles shown on the bottom panel of Fig.~\ref{fig:IVN_HD104237}
($\varphi_{\textrm{orb}}=0.934-0.950$) represent the situation where both
components are well separated. These observations were obtained within 7.6\,h.
We detect a definite magnetic field of positive polarity in both components.
The magnetic field of the primary is weak varying from 47\,G to 72\,G, in
agreement with previously reported values. Within the same time interval, the
magnetic field of the T\,Tauri component shows strong variability, decreasing
from $\bz =609\pm27$\,G to $\bz =124\pm13$\,G. 

\begin{figure}
 \centering 
        \includegraphics[width=\columnwidth]{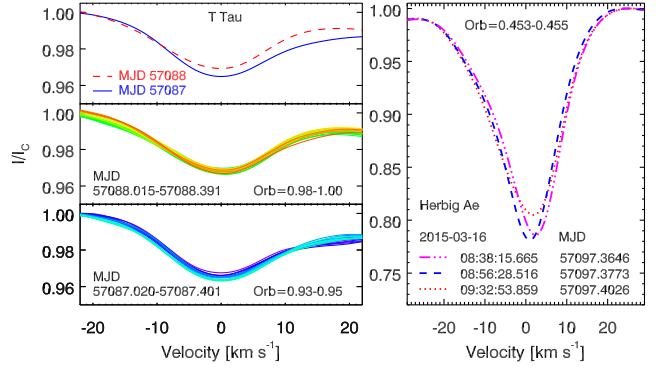}
        \caption{
          Line profile variability observed in the LSD Stokes~$I$ profiles. The
          profiles were shifted according to the orbital motion.
          {\em Left:}  Profiles of the secondary component of HD\,104237
          obtained during the nights when the components are well separated.
          There are 17 spectra taken within a nine hour interval during the
          night of MJD\,57087 (bottom panel) and 20 for the following night
          (MJD\,57088; middle panel). The total exposure time of each spectrum
          obtained from the combination of the four subexposures is about
          18~minutes. The average profiles for each night are shown on the top
          panel.
          {\em Right:} Profiles of the primary component during the night
          where we detect strong variability in the magnetic field strength.
        }
   \label{fig:Izoom}
\end{figure}

To measure longitudinal magnetic fields in other phases, where the components
are blended, we assumed that the contribution of the primary to the magnetic
field is weak and the strong Stokes~$V$ Zeeman features are related to the
T\,Tauri star. To be able to estimate its magnetic field strength, we have
calculated the average Stokes~$I$ profile of the secondary using the
individual profiles in spectra obtained in the phases where the components are
well separated. As we show in the left of Fig.~\ref{fig:Izoom}, the line
profile variability for the secondary component on short and long-time scales
is not very strong. 

The three SVD profiles shown in the middle panel of Fig.~\ref{fig:IVN_HD104237}
($\varphi_{\textrm{orb}}=0.453-0.455$) correspond to observations taken within
57 minutes. Within this time interval, the $\bz$ increases from 144\,G to
410\,G and then decreases back to 189\,G. As we show in the right of
Fig.~\ref{fig:Izoom}, also the Stokes~I profiles show strong variability at
this orbital phase range. The Stokes~$I$ and $V$ profiles shown in the top
panel of Fig.~\ref{fig:IVN_HD104237} ($\varphi_{\textrm{orb}}=0.544-0.547$) also
show significant variability, but the field strength remains almost constant
with $\bz\approx350$\,G over a 1.4\,h time interval.

\section{Searching for the rotation period of the primary}
\label{sec:Prot}

\begin{figure}
 \centering 
 \includegraphics[width=\columnwidth]{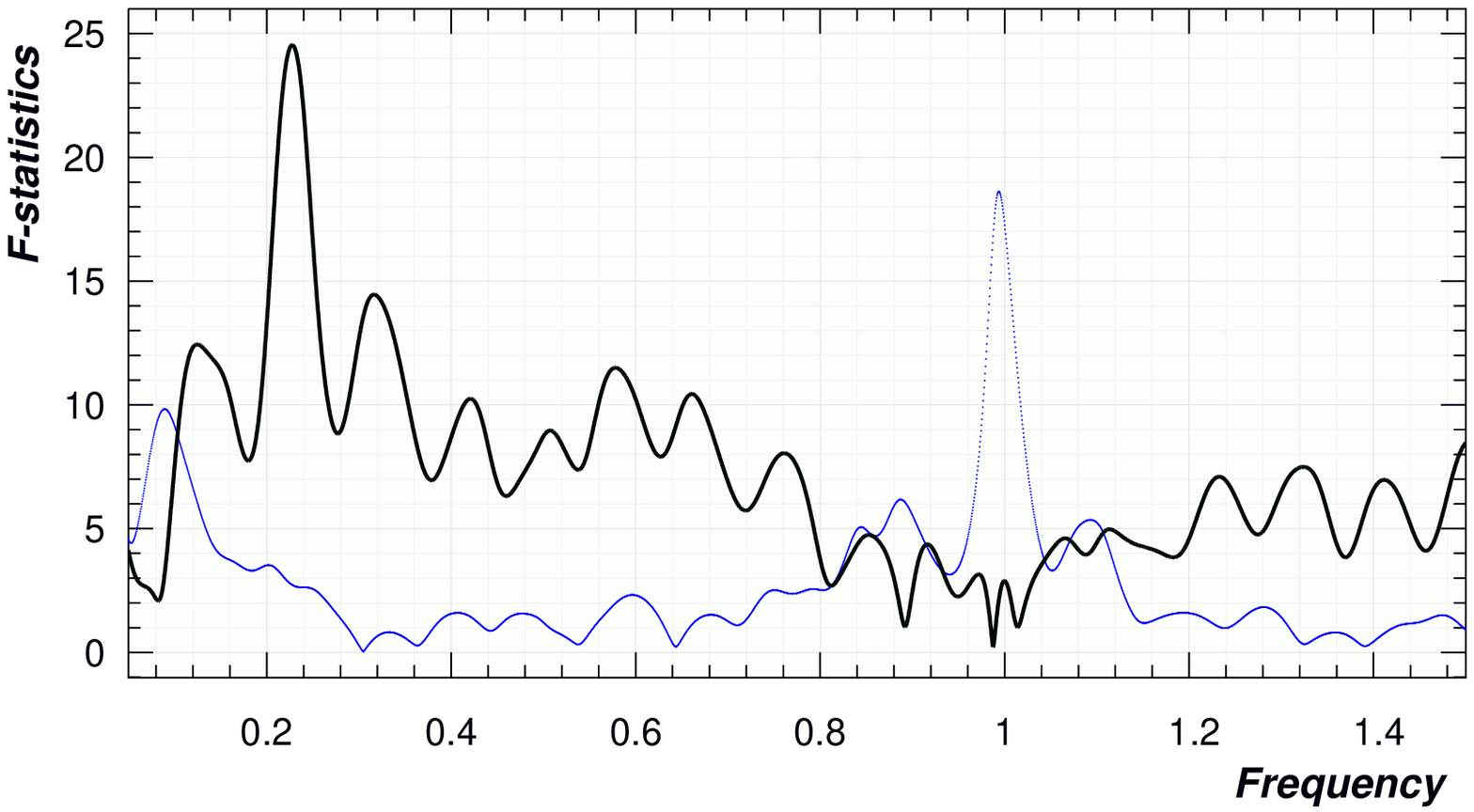}
 \includegraphics[width=\columnwidth]{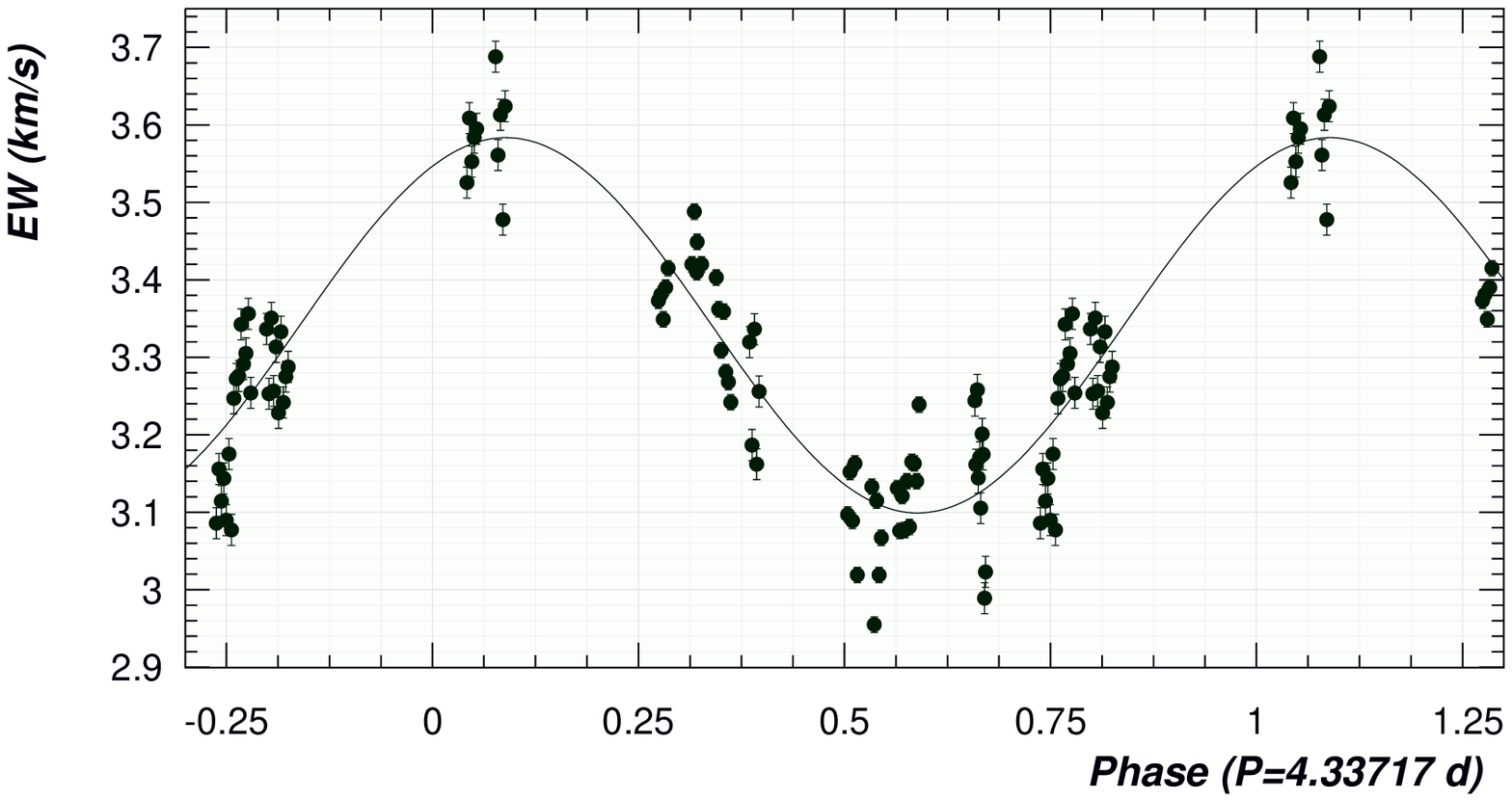}
        \caption{
          {\em Top:} Periodogram for the equivalent width measurements of the
          primary. The peak at 0.23057\,d$^{-1}$ corresponds to a period
          of\,4.33717\,d. The window function is denoted with a blue dotted
          line.
          {\em Bottom:} EWs measured in the primary phased with the period of
          4.33717\,d.
        }
   \label{fig:fstat}
\end{figure}

The EWs of the Stokes~$I$ LSD profiles of the primary component can be used to
search for a periodicity corresponding to the rotation period. The search for
the period was carried out using a non-linear least-squares fit to multiple
harmonics using the Levenberg--Marquardt method
\citep{press}.
We calculate the frequency spectrum with a specific number of trial frequencies
within the region of interest. A weighted linear least-squares fit is used for
each frequency to fit a sine curve and bias offset. Based on the result of the
fit, we make a statistical test to check the null hypothesis on the absence of
periodicity, i.e.\ to check the statistical significance of the amplitude of
the fit
\citep{seber}.
The resulting F-statistics presented in Fig.~\ref{fig:fstat} can be thought
of as the total sum, including covariances of the ratio of harmonic amplitudes
to their standard deviations. The measured EWs phased with the best period
of $4.33717\pm0.00316$\,d ($T_{0}=2\,457\,086.3$) are presented in
Fig.~\ref{fig:fstat}. The contribution of the secondary to the EWs
(0.36\,\kms) at phases where the components are blended was removed from the
measured values. The dispersion seen in the EW measurements is probably caused
by $\delta$\,Scuti-like pulsations in the primary component and a possible
presence of temperature spots in the secondary. Due to the weakness of the
Stokes~$I$ LSD profiles calculated for the secondary component, no analysis of
the its periodicity can be carried out.

\section{Discussion}
\label{sec:disc}

Our analysis of HARPSpol observations of the SB2 system HD\,104237 shows
that both components, the Herbig~Ae star and the lower mass T\,Tauri star
possess a magnetic field. The magnetic field of the Herbig\,Ae primary was
measured at phases where both components are well separated. In other phases,
where the components are blended, the Stokes~$V$ profiles were assumed to be
dominated by the magnetic field of the T\,Tauri secondary. The longitudinal
magnetic field of the Herbig~Ae star is weak and is only slightly changing
from 47\,G to 72\,G.
According to
\citet{alecian2014},
the magnetic properties of A- and B-type stars must have been shaped before the
Herbig~Ae/Be phase of stellar evolution. Using pre-main sequence evolutionary
tracks calculated with the CESAM code
\citep{morel},
she concluded that even stars above three solar masses will undergo a purely
convective phase before reaching the birthline. Therefore, it is plausible that
the weak magnetic fields detected in a number of Herbig~Ae/Be stars are just
leftovers of the fields generated by a dynamo mechanism during the convective
phase. If this scenario is valid, we should expect a significantly larger
number of Herbig stars possessing weak magnetic fields.

The rather strong longitudinal magnetic field of the T\,Tauri star was
estimated on multiple epochs and shows strong variability. In orbital phases
where the T\,Tauri component in the composite spectrum is well separated from
the Herbig~Ae star, $\left<B_{\rm z}\right>$ decreases from 609\,G to 124\,G
within 7.6\,h. The variability of EWs of the primary component indicates a
rotation period of $4.33717\pm0.00316$\,d ($104\pm0.08$\,hours), which is in
agreement with the value ($100\pm5$\,h) reported by
\citet{boehm2006},
based on measurements in H$\alpha$. No rotation period was determined for the
secondary due to its low flux contribution in the composite spectrum.

A search for magnetic fields and the determination of their geometries in
close binary systems is very important as the knowledge of the presence of a
magnetic field and of the alignment of the magnetic axis with respect to the
orbital radius vector in Herbig binaries may hint at the mechanism of the
magnetic field generation.
\citet{aksco}
reported a magnetic field detection on the secondary component of the
Herbig\,Ae double-lined spectroscopic binary AK\,Sco in the region of the
stellar surface facing permanently the primary component. This indicates that
the magnetic field geometry in the secondary component is likely related to
the position of the primary component. A similar magnetic field behaviour,
where the field orientation is linked to the companion, has previously been
detected in two other binaries, the so far only known close main-sequence
binaries with Ap components, HD\,98088 and HD\,161701
\citep{Babcock, Hubrig2014}.


\section*{Acknowledgments}

Based on observations made with ESO Telescopes at the La Silla Paranal
Observatory under programme IDs 085.D-0296 and 094.D-0704.
This work has made use of the VALD database, operated at Uppsala
University, the Institute of Astronomy RAS in Moscow, and the University of
Vienna.
N.A.D.\ acknowledges Russian Foundation for Basic Research (RFBR) according to
the research project 18-02-00554.
M.A.P.\ thanks the RFBR for support under grant 18-52-06004.





\begin{thebibliography}{99}

\bibitem[\protect\citeauthoryear{Alecian et al.}{2009}]{alecian2009}
Alecian E., et al.,
2009, \mnras, 400, 354

\bibitem[\protect\citeauthoryear{Alecian}{2014}]{alecian2014}
  Alecian E.,
  2014, in Putting A Stars into Context: Evolution, Environment, and Related
  Stars, eds.\ G.~Mathys, E.~R.~Griffin, O.~Kochukhov, R.~Monier,
  G.~M.~Wahlgren, p.\ 84

 \bibitem[\protect\citeauthoryear{Babcock}{1958}]{Babcock}
  Babcock H.~W.,
  1958, \apjs, 3, 141

 \bibitem[\protect\citeauthoryear{B\"ohm et al.}{2004}]{boehm2004}
 B\"ohm T., Catala C., Balona L., Carter B.,
 2004, \aap, 427, 907

 \bibitem[\protect\citeauthoryear{B\"ohm, Dupret \& Aynedjian}{2006}]{boehm2006}
 B\"ohm T., Dupret M.~A., Aynedjian H.,
 2006, \memsai, 77, 362

\bibitem[\protect\citeauthoryear{Carroll et al.}{2012}]{Carroll2012}
Carroll T.~A., Strassmeier K.~G., Rice J.~B., K\"unstler, A.,
2012, \aap, 548, A95

 \bibitem[\protect\citeauthoryear{Cowley, Castelli \& Hubrig}{2013}]{Cowley2013}
 Cowley C.~R., Castelli F., Hubrig S.,
 2013, \mnras, 431, 3485

\bibitem[\protect\citeauthoryear{Donati et al.}{1997}]{Donati1997}
Donati J.-F., Semel M., Carter B.~D., Rees D.~E., Collier Cameron A.,
1997, \mnras, 291, 658

\bibitem[\protect\citeauthoryear{Fumel \& B\"{o}hm}{2012}]{Fumel2012}
Fumel A., B\"{o}hm T., 2012,
A\&A, 540, 108

\bibitem[\protect\citeauthoryear{Garcia et al.}{2013}]{garcia2013}
Garcia P.~J.~V., et al.,
2013, \mnras, 430, 1839

\bibitem[\protect\citeauthoryear{Grady et al.}{2004}]{Grady2004}
Grady C.~A., et al., 2004,
ApJ, 608, 809

\bibitem[\protect\citeauthoryear{Hubrig et al.}{2011}]{Hubrig2011}
Hubrig S., et al.,
2011, A\&A, 525, L4

\bibitem[\protect\citeauthoryear{Hubrig et al.}{2013}]{Hubrig2013}
Hubrig S., Ilyin I., Sch{\"o}ller M., Lo Curto G.,
2013, Astron.\ Nachr., 334, 1093

\bibitem[\protect\citeauthoryear{Hubrig et al.}{2014}]{Hubrig2014}
Hubrig S., et al.,
2014, \mnras, 440, L6

\bibitem[\protect\citeauthoryear{Hubrig et al.}{2015}]{Hubrig2015}
Hubrig S., Carroll T.~A., Sch{\"o}ller M., Ilyin I.,
2015, \mnras, 449, L118

\bibitem[\protect\citeauthoryear{J\"arvinen et al.}{2015}]{silva2015}
J\"arvinen S.~P., Carroll T.~A., Hubrig S., Sch\"oller M., Ilyin I., Korhonen H., Pogodin M., Drake N.~A.,
2015, \aap, 584, 15

\bibitem[\protect\citeauthoryear{J\"arvinen et al.}{2018}]{aksco}
J\"arvinen S.~P., et al.,
2018, \apj, 858, L18

\bibitem[\protect\citeauthoryear{Kupka et al.}{2011}]{Kupka2011}
Kupka F., Dubernet M.-L., VAMDC Collaboration,
2011, Baltic Astronomy, 20, 503

\bibitem[\protect\citeauthoryear{Mathys}{1989}]{Mathys1989}
Mathys G.,
1989, Fundam.~Cosm.~Phys., 13, 143

\bibitem[\protect\citeauthoryear{McKee \& Ostriker}{2007}]{McKee}
  McKee C.~F., Ostriker E.~C.,
  2007, \araa, 45, 565

\bibitem[\protect\citeauthoryear{Morel}{1997}]{morel}
  Morel P.,
  1997, \aaps, 124, 597
  
\bibitem[\protect\citeauthoryear{Press et al.}{1992}]{press}
Press W.~H., Teukolsky S.~A., Vetterling W.~T., Flannery B.~P.,
1992, Numerical Recipes, 2nd edn.\ Cambridge: Cambridge University Press

\bibitem[\protect\citeauthoryear{Ryabchikova et al.}{2015}]{VALD3}
Ryabchikova T., Piskunov N., Kurucz R.~L., Stempels, H.~C., Heiter,
U., Pakhomov, Y., Barklem, P.~S.,
2015, \physscr, 90, 054005

\bibitem[\protect\citeauthoryear{Seber}{1977}]{seber}
Seber G.~A.~F.,
1977, Linear Regression Analysis.\ Wiley, New York

\bibitem[\protect\citeauthoryear{Snik et al.}{2008}]{snik2008}
Snik F., Jeffers S., Keller C., Piskunov N., Kochukhov O., Valenti J., Johns-Krull C.,
2008, SPIE Conf.\ Series, 7014, E22

\bibitem[\protect\citeauthoryear{Wade et al.}{2007}]{Wade2007}
Wade G.~A., Bagnulo S., Drouin D., Landstreet J.~D., Monin D.,
2007, \mnras, 376, 1145

\bibitem[\protect\citeauthoryear{Wade et al.}{2011}]{Wade2011}
Wade G.~A., Alecian E., Grunhut J., Catala C., Bagnulo S., Folsom C.~P., Landstreet J.~D.,
2011, in Proc.\ Conference on Astronomical Polarimetry 2008: Science from Small to Large Telescopes, eds.\ P.\ Bastien, N.\ Manset, D.~P.\ Clemens, \& N.\ St-Louis, ASP Conf.\ Ser., 449, 262 (San Francisco)

\end{thebibliography}




\bsp	
\label{lastpage}
\end{document}